\documentstyle{article}
\input epsf
\newfont{\SET}{msbm10 scaled \magstep3}
\newfont{\Set}{msbm10 scaled \magstep2}
\newfont{\set}{msbm10}
\newcommand{\p}{\mbox{\boldmath $\phi$}}
\newcommand{\s}{\mbox{\boldmath $\sigma$}}
\newcommand{\ta}{\mbox{\boldmath $\tau$}}
\newcommand{\la}{\mbox{\boldmath $\lambda$}}
\newcommand{\M}{\mbox{\boldmath ${\rm M}$}}
\newcommand{\CP}{$\mbox{\set C}P^{1}$\ }
\newcommand{\R}{$\mbox{\set R}^{2}$}
\newcommand{\Ssp}{$S_{\scriptscriptstyle {\rm sp}}^{2}$}
\newcommand{\Siso}{$S_{\rm {\scriptscriptstyle iso}}^{2}$}

\title{Low-energy dynamics of a $\mbox{\SET C}P^{1}$ lump on the sphere}
\author{J M Speight\thanks{Present address: Department of Mathematics, 
University of Texas at Austin, Austin TX 78712, USA} \\
        Department of Mathematical Sciences, \\
        University of Durham, Durham DH1 3LE, UK}
\date{}

\begin{document}
\maketitle

\begin{abstract}

Low-energy dynamics in the unit-charge sector of the $\mbox{\set C}P^{1}$ model
on spherical space (space-time $S^{2}\times\mbox{\set R}$) is treated in the 
approximation of geodesic motion on the moduli space of static solutions, a 
six-dimensional manifold with non-trivial topology and metric. The structure of 
the induced metric is restricted by consideration of the isometry group 
inherited from global symmetries of the full field theory. Evaluation of the 
metric is then reduced to finding five functions of one coordinate, which may  
be done explicitly. Some totally geodesic submanifolds are found and the 
qualitative features of motion on these described.

\end{abstract}

\section{Introduction}

The $\mbox{\set C}P^{1}$ model in flat space is a scalar field theory whose
configuration space $Q$ consists of finite energy maps from Euclidean \R\
to the complex projective space $\mbox{\set C}P^{1}$, the energy
functional being constructed naturally from the Riemannian structures of the
base and target spaces (that is, the model is a pure sigma model in the broad
sense). The requirement of finite energy imposes a boundary condition at spatial
infinity, that the field approaches the same constant value, independent of
direction in \R, so that the field may be regarded as a map from the one point
compactification $\mbox{\set R}^{2}\cup\{\infty\}\cong S^{2}$ to \CP. Since
$\mbox{\set C}P^{1}\cong S^{2}$ also, finite energy configurations are effectively 
maps $S^{2} \rightarrow S^{2}$, the homotopy theory of which is well understood, and the
configuration space is seen to consist of disconnnected sectors $Q_{n}$ labelled 
by an integer $n$, the
``topological charge'' (degree),
\begin{equation}
Q=\bigcup_{n\in \mbox{\set Z}}Q_{n}.
\end{equation}
 Each configuration is trapped within its
own sector because time evolution is continuous.

The Lorentz invariant, time-dependent model is not integrable but complete 
solution of the static problem has been
achieved by means of a Bogomol'nyi argument and the general charge $n$ moduli
space, the space of charge-$n$ static solutions $\M_{n}\subset Q_{n}$, is known
(that {\em all}\, static, finite energy solutions of the \CP model saturate the
Bogomol'nyi bound is a non-trivial result \cite{Din}). 
Each static solution within the charge-$n$ sector has the same
energy (minimum within that sector and proportional to $n$), and $\M_{n}$ is
parametrized by $4n+2$ parameters (the moduli), so such a moduli
space may be thought of as the $(4n+2)$-dimensional level bottom of a potential
valley defined on the infinite dimensional charge-$n$ sector, $Q_{n}$. Low energy
{\em dynamics} may be approximated by motion restricted to this valley bottom,
a manifold embedded in the full configuration space, and thus inheriting from it
a non-trivial metric induced by the kinetic energy functional. The approximate
dynamic problem is reduced to the geodesic problem with this metric, and has
been investigated by several authors \cite{Ward,Leese}. In the unit-charge sector one here
encounters a difficulty: certain components of the metric are singular and the
approximation is ill defined. For example, unit-charge static solutions are
localized lumps of energy with arbitrary spatial scale, so one of the six moduli
of $\M_{1}$ is a scale parameter. Motion which changes this parameter is
impeded by infinite inertia in the geodesic approximation, a result in conflict
with numerical evidence which suggests that lumps collapse under scaling
perturbation \cite{Zak}.

This problem should not be present in the model defined on a compact two
dimensional physical space. The obvious choice is the $2$-sphere because the
homotopic partition of the configuration space carries through unchanged. Also,
$S^{2}$ with the standard metric is conformally equivalent to Euclidean
$\mbox{\set R}^{2}\cup\{\infty\}$, and the static $\mbox{\set C}P^{1}$ model energy functional
is conformally invariant, so the whole flat space static analysis is still
valid and all the moduli spaces are known. However, the kinetic energy functional
{\em does} change and induces a new, well defined metric on the unit-charge 
moduli space. By means of the isometry group derived from the spatial and
internal symmetries of the full field theory we can place restrictions on the
possible structure of this metric, greatly simplifying its evaluation. The
geodesic problem is still too complicated to be solved analytically in general,
but by identifying totally geodesic submanifolds, it is possible to obtain the
qualitative features of a number of interesting solutions. In particular, the
possibilities for lumps travelling around the sphere are found to be
unexpectedly varied.

\section{The $\mbox{\Set C}P^{1}$ model on $S^{2}$}

The $\mbox{\set C}P^{1}$ model on the $2$-sphere is defined by the Lagrangian
\begin{equation}
L[W]=\int_{S^{2}}\, dS\, \frac{\partial_{\mu}W\partial_{\nu}\bar{W}}{(1+|W|^{2})^{2}}\, 
     \eta^{\mu\nu}
\end{equation}
where $W$ is a complex valued field, $dS$ is the invariant $S^{2}$ measure and
$\eta^{\mu\nu}$ are the components of the inverse of the Lorentzian metric
\begin{equation}
\eta = dt^{2}-d^{2}\Omega
\end{equation}
on {\set R}(time)$\times S^{2}$(space), $d^{2}\Omega$ being the natural metric
on $S^{2}$. Although the 
language of the $\mbox{\set C}P^{1}$ model is analytically convenient, the
homotopic classification and physical meaning of the field configurations are
more easily visualized if we exploit the well known equivalence to the
$O(3)$ sigma model \cite{Woo,Belavin}. In the latter, the scalar field is a three dimensional 
isovector \p\  constrained to have unit length with respect to the Euclidean
$\mbox{\set R}^{3}$ norm ($\p \cdot \p \equiv 1$), that is, the target space is
the 2-sphere of unit radius with its natural metric, which we will denote \Siso\ 
for clarity. (The suffix refers to ``isospace'' in analogy with the internal
space of nuclear physics models.) The \CP field $W$ is then thought of as the
stereographic image of \p\  in the equatorial plane, projected from the North
pole, $(0,0,1)$. Explicitly,
\begin{equation}
\label{eq:phidef}
\p =\left(\frac{W+\bar{W}}{1+|W|^{2}}, \frac{W-\bar{W}}{i(1+|W|^{2})},
          \frac{|W|^{2}-1}{1+|W|^{2}}\right)
\end{equation}
and
\begin{equation}
W=\frac{\phi_{1}+i\phi_{2}}{1-\phi_{3}}.
\end{equation}
Then
\begin{equation}
L[W]\equiv L_{\sigma}[\p]=\frac{1}{4}\int_{S^{2}}\, dS\, 
        \partial_{\mu}\p\cdot\partial_{\nu}\p\, \eta^{\mu\nu}
\end{equation}
the familiar $O(3)$ sigma model Lagrangian. A $W$ configuration, then, may be
visualized as a distribution of unit length arrows over the surface of the 
physical 2-sphere \Ssp.  
Each smooth map \Ssp $\rightarrow$\Siso\ falls into one of a discrete infinity
of disjoint homotopy classes, each class associated with a unique integer which
may be thought of as the topological degree of the map 
(see, for example \cite{Nakahara}),
 so homotopic partition of the 
configuration space is built in to the model from the start.    

We also choose stereographic coordinates $(x,y)$ on \Ssp, in terms of
which,
\begin{equation}
(\eta_{\mu\nu})={\rm diag}\left(1,\frac{-1}{(1+r^{2})^{2}},\frac{-1}{(1+r^{2})^{2}}
                          \right)
\end{equation}
where $r=\sqrt{x^{2}+y^{2}}$, $(x,y)$ takes all values in $\mbox{\set R}^{2}$
and $x^{0}=t$, $x^{1}=x$, $x^{2}=y$. The radius of \Ssp\ has been normalized to 
unity. The invariant measure is,
\begin{equation}
dS=dx\, dy\, \sqrt{|\det(\eta_{\mu\nu})|}=\frac{dx\, dy}{(1+r^{2})^{2}}
\end{equation}
and so,
\begin{equation}
L[W] = \int\, dx\, dy\, \frac{1}{(1+|W|^{2})^{2}}\left(
\frac{|\dot{W}|^{2}}{(1+r^{2})^{2}}-\left|\frac{\partial W}{\partial x}\right|^{2}
                                   -\left|\frac{\partial W}{\partial y}\right|^{2}
                                   \right).  
\end{equation}
We identify kinetic energy,
\begin{equation}
T[W] = \int\, \frac{dx\, dy}{(1+r^{2})^{2}}\frac{|\dot{W}|^{2}}{(1+|W|^{2})^{2}}
\end{equation}
and potential energy
\begin{equation}
V[W] = \int\, dx\, dy\, \frac{1}{(1+|W|^{2})^{2}}\left(
                          \left|\frac{\partial W}{\partial x}\right|^{2}+
                           \left|\frac{\partial W}{\partial y}\right|^{2}
                                   \right).  
\end{equation}
Note that the potential energy is identical to that for flat space by virtue of 
the conformal invariance of the static model (stereographic projection is a
conformal transformation). Thus the familiar Bogomol'nyi argument \cite{Rajaraman}
follows immediately and ($i,j,k$ run over $1,2$ and $\times$ represents the
$\mbox{\set R}^{3}$ vector product in \p\ space):
\begin{eqnarray}
0 &\leq& \int\, dx\, dy\, 
\left(\partial_{i}\p \pm \epsilon_{ij}\p\times\partial_{j}\p\right)\cdot
\left(\partial_{i}\p \pm \epsilon_{ik}\p\times\partial_{k}\p\right) \nonumber \\
&=& 2\int\, dx\, dy\, 
\left[\partial_{i}\p\cdot\partial_{i}\p\mp\epsilon_{ij}
(\partial_{i}\p\times\partial_{j}\p)\cdot\p\right], \nonumber \\
\Rightarrow V[W] &=& \frac{1}{4}\int\, dx\, dy\, 
\partial_{i}\p\cdot\partial_{i}\p \nonumber \\
&\geq& \frac{1}{2}\left|\int\, dx\, dy\, 
\left(\frac{\partial\p}{\partial x}\times\frac{\partial\p}{\partial y}\right)\cdot\p\right|
=2\pi|n|,
\end{eqnarray}
where $W$ is in the degree $n$ homotopy class, equality holding if and only if
\begin{equation}
\partial_{i}\p=\mp\epsilon_{ij}\p\times\partial_{j}\p,
\end{equation}
which, on substitution of (\ref{eq:phidef}) becomes the Cauchy-Riemann condition
for $W$ to be an analytic function of $z=x+iy$ (upper sign) or $\bar{z}=x-iy$
(lower sign). The former (latter) case corresponds to static solutions of
positive (negative) degree, and if $W$ is single valued with finite degree $n$,
then it must be a rational map of degree $n$ in $z$ if $n\geq 0$ or in $\bar{z}$
if $n<0$. We shall deal with the unit charge moduli space, consisting of all
rational maps of degree 1 in $z$. Since the configuration space and moduli
spaces of the flat space and spherical space models are diffeomorphic, we shall
use the same notation ($Q,Q_{n},\M_{n}$ etc.) in both cases.                                  
        
\section{The unit-charge moduli space}

The simplest static unit-charge solution is 
\begin{equation}
W(z)=z
\end{equation}
which we shall call the symmetric hedgehog because its \p\  field points radially 
outwards at all points on \Ssp. Its energy density is uniformly 
distributed, so it is not really a lump. Since the static model is conformally invariant,
any configuration obtained from this by a M\"{o}bius transformation must be
another point on the moduli space. In fact the orbit of $W=z$ under the 
M\"{o}bius group
{\em is} the space of degree 1 rational maps, each map being generated by one
and only one group element. Thus we may identify the moduli space with the
parameter space of the M\"{o}bius group.

There is a well known matrix representation of M\"{o}bius transformations 
\cite{Schwer} which we denote thus:
\begin{equation}
\label{eq:dotdef}
W(z)=\frac{az+b}{cz+d}=\left(\begin{array}{cc}a&b\\c&d\end{array}\right)\odot z
=M \odot z
\end{equation}
where $M\in GL(2,\mbox{\set C})$ so that $\det M \neq 0$. The last condition 
ensures the
invertibilty of the transformation and fixes the degree of $W$ at 1. The 
M\"{o}bius group product becomes matrix multiplication,
\begin{equation}
M_{2}\odot(M_{1}\odot z)=(M_{2}M_{1})\odot z
\end{equation}
where the left hand side means
\begin{equation}
M_{2}\odot(M_{1}\odot z)=\frac{a_{2}(M_{1}\odot z)+b_{2}}{c_{2}(M_{1}\odot z)+d_{2}}
\end{equation}
in obvious notation. All matrices differing by a constant factor yield the same
configuration, and $\det M \neq 0$ so when we divide by this scaling equivalence
we can choose a unimodular matrix as the representative for each equivalence class.
There are two such matrices possible for each distinct configuration because if
$M$ is unimodular, so is $-M$. Thus $SL(2,\mbox{\set C})$ is a double cover of
the moduli space, which we recover by dividing out the equivalence $M'\sim M
\Leftrightarrow M'=-M$: the moduli space is $SL(2,\mbox{\set C})/\mbox{\set Z}_{2}$.

Coincidentally, $SL(2,\mbox{\set C})$ is also a double cover of the proper orthochronous
Lorentz group. The statement that any Lorentz transformation may be formed by a
unique composition of a boost then a rotation (or {\em vice versa}) translates to the
existence, for all $M\in SL(2,\mbox{\set C})$, of $U\in SU(2)$ and $H$, a
positive definite, unimodular, Hermitian $2\times2$ matrix (call this set
${\cal H}$), satisfying
\begin{equation}
M=UH
\end{equation}
both $U$ and $H$ being unique \cite{Penrose}. It follows that the space 
$SL(2,\mbox{\set C})$ is locally a product of $S^{3}$ (the group manifold 
of $SU(2)$) and $\mbox{\set R}^{3}$ (the parameter space of ${\cal H}$), a result
which generalizes globally, $SL(2,\mbox{\set C})\cong S^{3}\times\mbox{\set R}^{3}$.

We may choose local coordinates on $SL(2,\mbox{\set C})$ by defining the standard
Euler angles $(\alpha,\beta,\gamma)$ on $S^{3}$,
\begin{equation}
U=\left(\begin{array}{cc} \cos\frac{\alpha}{2}e^{i(\beta+\gamma)/2} &
                          \sin\frac{\alpha}{2}e^{i(\beta-\gamma)/2} \\
                          -\sin\frac{\alpha}{2}e^{-i(\beta-\gamma)/2} &
                          \cos\frac{\alpha}{2}e^{-i(\beta+\gamma)/2}
                          \end{array} \right)
\end{equation}
and expanding $H$ in terms of Pauli matrices \ta,
\begin{equation}
H=\Lambda\mbox{\set I}+ \la\cdot\ta 
=\left(\begin{array}{cc} \Lambda+\lambda_{3} & \lambda_{1}-i\lambda_{2} \\
                         \lambda_{1}+i\lambda_{2} & \Lambda-\lambda_{3}
                         \end{array} \right)
\end{equation}
$\Lambda(\la)$ being chosen  to ensure the unimodular and postive definite
properties:
\begin{equation}
\Lambda = \sqrt{1+\lambda^{2}}.
\end{equation}
The 3-vector \la\ (modulus $|\la|=\lambda$) takes all values in $\mbox{\set R}^{3}$,
while $\beta\in [0,4\pi]$, $\gamma \in [0,2\pi]$ and $\alpha \in [0,\pi]$. These 
ranges allow $M$ to take all values in the double cover $SL(2,\mbox{\set C})$.
In analyzing the structure of the metric, it is convenient to work with
$SL(2,\mbox{\set C})$, checking that the metric is single valued under the 
identification of $M$ with $-M$. The true moduli space
$SL(2,\mbox{\set C})/\mbox{\set Z}_{2}$ is charted by the same coordinates but
with $\beta$ lying in the reduced range $[0,2\pi]$, for $U$ is then restricted
to the ``upper half'' of $S^{3}$. The
chart has a coordinate singularity at $\alpha=0$ and at $\alpha=\pi$. The explicit
connexion between a point in $\M_{1}$ and the corresponding static solution will
be made in section 5, below.

\section{The induced metric and its isometry group}

Field dynamics of the \CP model may be visualized as the dynamics of a point
particle with ``position'' $W:S^{2}_{\rm {\scriptscriptstyle sp}}\rightarrow 
S^{2}_{\rm {\scriptscriptstyle iso}}$ moving in an infinite-dimensional 
configuration space. A solution $W(t,x,y)$ of the field equations is thought of
as a trajectory in this space, motion on which is determined by metric $T[W]$
and potential $V[W]$. In the unit-charge sector, the Bogomol'nyi argument shows
that there is a six-dimensional subspace on which the potential achieves its
topological minimum value of $2\pi$, and that any perturbation departing from
this subspace must involve increasing $V$. If a configuration sitting at the 
bottom of this potential valley is given a small velocity tangential to it then
we expect the ensuing time-evolved field to stay close to the valley bottom,
for departure from it entails climbing up the valley walls. In the geodesic
approximation \cite{Manton} we restrict motion to the valley bottom, assuming
that orthogonal modes are insignificant.

Thus, at all times $W(t,x,y)$ is a solution of the {\em static model}, but we
allow the moduli $\{\la,\alpha,\beta,\gamma\}$, denoted collectively by
$\{q^{i}: i=1 \ldots 6\}$, to vary with time in accordance with the inherited
action principle. So,
\begin{equation}
\dot{W}=\frac{\partial W}{\partial q^{i}} \dot{q}^{i},
\end{equation}
and the Lagrangian is
\begin{equation}
L=T-V=\int\, \frac{dx\, dy}{(1+r^{2})^{2}}\frac{\partial W}{\partial q^{i}}
\frac{\partial \bar{W}}{\partial q^{j}}\dot{q}^{i}\dot{q}^{j}\, -2\pi.
\end{equation}
Defining the induced metric,
\begin{equation}
g(q)=g_{ij}(q)\, dq^{i}\, dq^{j}=2\int\, \frac{dx\, dy}{(1+r^{2})^{2}}\frac{\partial W}{\partial q^{i}}
\frac{\partial \bar{W}}{\partial q^{j}}\, dq^{i}\, dq^{j}
\label{eq:metricdef}
\end{equation}
and ignoring the irrelevant constant, the Lagrangian is recast as that of a free
particle moving on a Riemannian manifold with metric $g$:
\begin{equation}
L=\frac{1}{2}g_{ij}(q)\dot{q}^{i}\dot{q}^{j}.
\end{equation}
The equations of motion are the geodesic equations. In principle all we need do 
is evaluate the integrals of (\ref{eq:metricdef}), but these are 21 functions 
of 6 variables so as it stands this is
intractable in practice. It is profitable to take a more circumspect approach,
using symmetries of the model to place restrictions on the structure of $g$.

Consider the rotation group $SO(3)$ acting on \Ssp\ and \Siso. The former is the
group of spatial rotations (under which $W$, or equivalently \p, 
transforms as a scalar) while the latter is the group of global internal 
rotations (henceforth called ``isorotations'') of the \p\ field of which $W$ is 
the stereographic image.  Any such transformation ${\cal T}$ leaves invariant 
(a) the topological charge, so
${\cal T}$ is a bijection ${\cal T}:Q_{n}\rightarrow Q_{n}$,
(b) the potential energy, so within $Q_{n}$ static solutions are mapped to other
static solutions, ${\cal T}:\M_{n}\rightarrow\M_{n}$, and
(c) the kinetic energy, which induces the metric on $\M_{n}$. Hence ${\cal T}$
is an isometry of $(\M_{1},g)$. The $SU(2)$
subgroup of the M\"{o}bius group's double cover, $SL(2,\mbox{\set C})$, acting 
via the operation $\odot$ defined by equation (\ref{eq:dotdef}) is the double 
cover of the group of rotations of the 2-sphere \cite{Schwer} considered as 
operations on the  projective plane (spatial or internal, ie acting on $z=x+iy$ 
or $W$). Thus we find that ($L,R\in SU(2), M\in SL(2,\mbox{\set C})$),
\begin{equation}
M\mapsto LM \Rightarrow W(z)\mapsto L\odot(W(z))
\end{equation}
produces an isorotation of the configuration
$W(z)=M\odot z$, while
\begin{equation}
M\mapsto MR \Rightarrow W(z)\mapsto W(R\odot z)
\end{equation}
produces a spatial rotation, both isometries of the induced metric.

The action of the isorotation on the moduli space is simple:
\begin{eqnarray}
M=UH &\mapsto& LUH \\ \nonumber
\Rightarrow U &\mapsto& LU \\ \nonumber
H &\mapsto& H.
\end{eqnarray}
The isometry takes the $SU(2)$ left multiplication action on $S^{3}/\mbox{\set Z}_{2}$ while
leaving the $\mbox{\set R}^{3}$ moduli \la\ unchanged. Using a technique standard
in the analysis of isometries in general relativity \cite{Ryan}, we change
from the coordinate basis on $S^{3}/\mbox{\set Z}_{2}, \{d\alpha,d\beta,d\gamma\}$, to a 
non-coordinate basis, in this case the left-invariant 1-forms of the Lie group
$SU(2)$. These may be found by expanding the left-invariant 1-form
$U^{-1}dU$ in terms of a convenient basis of the Lie algebra $su(2)$, for
example $i\ta/2$. Explicitly,
\begin{equation}
U^{-1}dU=\s\cdot\left(\frac{i}{2}\ta\right)
\end{equation}
where
 \begin{eqnarray}
\label{eq:sigma}
\sigma_{1} &=& -\sin\gamma\, d\alpha +\cos\gamma\sin\alpha\, d\beta \\ \nonumber
\sigma_{2} &=&  \cos\gamma\, d\alpha +\sin\gamma\sin\alpha\, d\beta \\ \nonumber 
\sigma_{3} &=&  \cos\alpha\, d\beta +d\gamma.
\end{eqnarray}
If we evaluate the metric at one particular point on $S^{3}/\mbox{\set Z}_{2}$, for all possible
$\la\in\mbox{\set R}^{3}$, we can obtain the metric at all other points on
$S^{3}/\mbox{\set Z}_{2}$ because $S^{3}/\mbox{\set Z}_{2}$ is the isorotation orbit of our base point, and 
isorotation is an isometry, so the metric must remain constant (for each \la)
over the entire orbit. ``Constant'' means unchanging when considered as a
geometric object, not that the components with respect to the original
coordinate basis are constant, because the basis vectors themselves transform
non-trivially. The basis of (\ref{eq:sigma}) {\em is} invariant however, so the
metric must be of the form
\begin{equation}
g=\mu_{ab}(\la)\sigma_{a}\sigma_{b}+\nu_{ab}(\la)\sigma_{a}d\lambda_{b}
  +\pi_{ab}(\la)d\lambda_{a}d\lambda_{b},
\end{equation}
where $a,b=1,2,3$ and each of the component functions is independent of
$(\alpha,\beta,\gamma)$.  

Let us now consider the spatial rotations:
\begin{eqnarray}
M=UH &\mapsto& MR=UR\, R^{\dagger}HR \\ \nonumber
\Rightarrow U &\mapsto& UR \\ \nonumber
            H &\mapsto& R^{\dagger}HR \in {\cal H}.
\end{eqnarray}
The latter in terms of coordinates is
\begin{equation}
H=\sqrt{1+\lambda^{2}}\, \mbox{\set I} +\la\cdot\ta
\mapsto \sqrt{1+\lambda^{2}}\mbox{\set I} +R^{\dagger}(\la\cdot\ta)R.
\end{equation}
The action of conjugation of the Hermitian, traceless matrix $\la\cdot\ta$ by a
unitary matrix $R$ is well known \cite{Choquet} -- it is equivalent to a
$SO(3)$ rotation of \la:
\begin{equation}
R^{\dagger}(\la\cdot\ta)R=({\cal R}\la)\cdot\ta
\end{equation}
where ${\cal R}\in SO(3)$ with components ${\cal R}_{ab}=\frac{1}{2}{\rm tr}
(\tau_{a}R^{\dagger}\tau_{b}R)$. The action on the left-invariant 1-forms \s\ is
similar. Under $U\mapsto UR$,
\begin{eqnarray}
U^{-1}dU &\mapsto& R^{\dagger}(U^{-1}dU)R \\ \nonumber
\Rightarrow \frac{i}{2}\s\cdot\ta &\mapsto& \frac{i}{2}R^{\dagger}(\s\cdot\ta)R=\frac{i}{2}({\cal R}\s\cdot\ta) \\ \nonumber
\Rightarrow \s &\mapsto& {\cal R}\s
\end{eqnarray}
where ${\cal R}$ is the same $SO(3)$ matrix defined above. Thus both \la\ and \s\
transform as 3-vectors under spatial rotations and as scalars under isorotations.

The metric must be invariant under spatial rotations also, so the task is to
construct from \la, $d\la$ and \s\, the most general possible $(0,2)$ tensor
which is scalar under these rotations. This is
\begin{eqnarray}
  \label{eq:metric}
g &=& A\, d\la\cdot d\la+B(\la\cdot d\la)^{2}+C\, \s\cdot\s+D(\la\cdot\s)^{2} \\ \nonumber
  & &  +E\, \s\cdot d\la +F(\la\cdot d\la)(\la\cdot\s)+G(\la\times\s)\cdot d\la
\end{eqnarray}
$A$---$G$ being 7 unknown functions of $\lambda=|\la|$ only.

The metric may be restricted still further on consideration of a discrete
isometry. The kinetic energy is invariant under the discrete ``parity''
transformations $P_{z}: z \mapsto \bar{z}$ and $P_{\scriptscriptstyle W}: W \mapsto \bar{W}$.
However, neither is an isometry of the moduli space because each reverses the
sign of the topological charge, mapping lumps to anti-lumps. The composite
transformation $P_{\scriptscriptstyle W}\circ P_{z}$ {\em is} an isometry. Using the configuration of
(\ref{eq:dotdef}),
\begin{equation}
W(z)\stackrel{P_{z}}\longmapsto\frac{a\bar{z}+b}{c\bar{z}+d}
\stackrel{P_{\scriptscriptstyle W}}\longmapsto\frac{\bar{a}z+\bar{b}}{\bar{c}z+\bar{d}}
=\bar{M}\odot z.
\end{equation}
In terms of the moduli, $M\mapsto\bar{M}$ is the transformation,
\begin{eqnarray}
\s &=& (\sigma_{1},\sigma_{2},\sigma_{3})\mapsto(-\sigma_{1},\sigma_{2},-\sigma_{3}) \\ \nonumber
\la&=& (\lambda_{1},\lambda_{2},\lambda_{3})\mapsto(\lambda_{1},-\lambda_{2},\lambda_{3}).
\end{eqnarray}
This isometry removes two of the terms in (\ref{eq:metric}) because under it,
\begin{equation}
\s\cdot d\la \mapsto -\s\cdot d\la,
\end{equation}
and
\begin{equation}
(\la\cdot d\la)(\la\cdot\s) \mapsto -(\la\cdot d\la)(\la\cdot\s)
\end{equation}
so that $E(\lambda)\equiv F(\lambda)\equiv 0$.

The remaining five functions of $\lambda$ are evaluated by choosing convenient
orientations for \la, positions on $S^{3}/\mbox{\set Z}_{2}$ and tangent vectors (velocities),
then calculating the kinetic energy and comparing with (\ref{eq:metric}).
Repeating this four times it is possible to extract the following (see figure 
1):
\begin{eqnarray}
A &=& 4\pi S_{2}(\chi) \\ \nonumber
B &=& \frac{4\pi}{\lambda^{2}}\left[\frac{4}{1+\lambda^{2}}S_{1}(\chi)-S_{2}(\chi)\right] \\ \nonumber
C &=& \frac{\pi}{2}-2\pi S_{1}(\chi) \\ \nonumber
D &=& \frac{\pi}{\lambda^{2}}\left[6S_{1}(\chi)-\frac{1}{2}\right] \\ \nonumber
G &\equiv& A
\end{eqnarray}
where,
\begin{eqnarray}
\chi(\lambda) &=& \frac{\sqrt{1+\lambda^{2}}+\lambda}{\sqrt{1+\lambda^{2}}-\lambda} \\ \nonumber
S_{1}(\chi)   &=& \frac{\chi^{2}}{2(\chi^{2}-1)^{3}}\left[(\chi^{2}+1)\log\chi^{2}-2\chi^{2}+2\right] \\ \nonumber
S_{2}(\chi)   &=& \frac{\chi}{(\chi^{2}-1)^{3}}\left[\chi^{4}-2\chi^{2}\log\chi^{2}-1\right]. 
\end{eqnarray}
Note that $\chi$ is a strictly increasing function of $\lambda$, and that
$\chi:[0,\infty)\rightarrow[1,\infty)$. There appear to be divergences of the
functions $A$---$D$ at $\lambda=0$, but these are in fact removable singularities,
so all the limits of vanishing $\lambda$ exist. Although $B$ and $D$ are negative
it is straightforward to show that this
metric is positive definite, as of course it must be. The veracity of the statement $G \equiv A$ is established by explicit calculation, there being no
obvious symmetry argument in its favour. In summary then, the
metric is
\begin{equation}
\label{eq:g}
g=A\, d\la\cdot d\la+B(\la\cdot d\la)^{2}+C\, \s\cdot\s+D(\la\cdot\s)^{2}+
  A(\la\times\s)\cdot d\la.
\end{equation}  

\vbox{
\centerline{\epsfysize=3truein
\epsfbox[70 190 550 590]{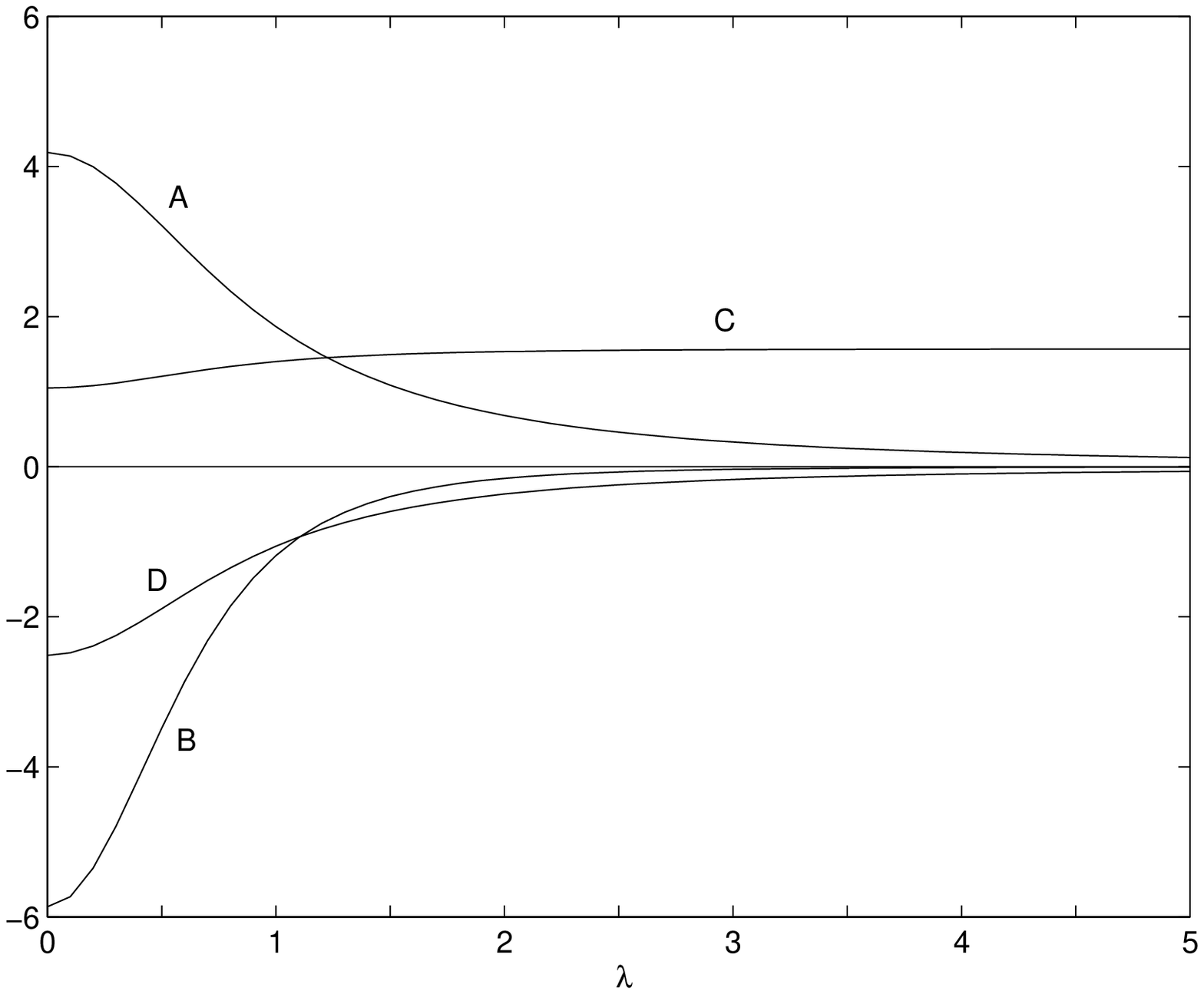}}
\centerline{\it Figure 1: The metric functions $A,B,C$ and $D$.}
}
\vspace{0.5cm}

\section{Some totally geodesic submanifolds}

Before discussing geodesics of the metric (\ref{eq:g}) we must describe
the connexion between a point on the moduli space $(S^{3}/\mbox{\set Z}_{2})\times\mbox{\set R}^{3}$
and its corresponding field configuration. Consider first the 3-dimensional 
submanifold defined by $U=\mbox{\set I}$, parametrized by $\la \in \mbox{\set R}^{3}$. Any
point in this subspace may be written as $\la={\cal R}\la'$ where
$\la'=(0,0,\lambda)$, $\lambda > 0$ and ${\cal R} \in SO(3)$. The lump represented
by $\la'$ is
\begin{equation}
\label{eq:hedgehog}
W'(z)=(\la'\cdot\ta)\odot z =
\left(\frac{\sqrt{1+\lambda^{2}}+\lambda}{\sqrt{1+\lambda^{2}}-\lambda}\right)z
=\chi(\lambda)z.
\end{equation}
This is a distorted hedgehog with the arrows pulled towards the North pole. The
larger $\lambda$ is, so the larger $\chi$ is and the greater is the distortion.
Although it is  usual to define the position of a lump as the position of  
maximum energy density, we shall refer to this as a lump of sharpness 
$\lambda$ located at $\widehat{\la}'=(0,0,1)$, the antipodal point to the energy density
peak which occurs where the arrows are stretched apart. Obviously the motion
of any point is trivially mirrored by its antipodal image, so this 
terminology makes sense. The lump represented by \la\ is
\begin{eqnarray}
W(z) &=& ({\cal R}\la'\cdot\ta)\odot z =[R^{\dagger}(\la'\cdot\ta)R]\odot z \\ \nonumber
&=& [R^{\dagger}(\la'\cdot\ta)]\odot(R\odot z)=R^{\dagger}\odot[W'(R\odot z)].
\end{eqnarray}
This configuration is formed by first performing a spatial rotation -- taking the
\p\ arrow at the old point $z$ and placing it at the new point $R\odot z$
without changing its orientation -- then performing the inverse isorotation. The
result looks like the arrows have been fixed to \Ssp\ which has then been
rotated by ${\cal R}$ which, as defined, has the action on $S^{2}$ equivalent to
$R^{\dagger}$, {\em not} $R$, acting on \CP\ via $\odot$. That is, if we define
${\rm P}$ to be stereographic projection, ${\rm P}:S^{2}\rightarrow$\CP so that
${\rm P}:\p\mapsto W$, then ${\rm P}:{\cal R}\p\mapsto R^{\dagger}\odot W$. So the
lump at the North pole is shifted to $\widehat{\la}={\cal R}\widehat{\la}'$.

All other points on the moduli space are on the isorotation orbit of this
submanifold, and isorotation, while changing the internal orientation of the
lump, does not move the lump around on physical space. Thus we can always
interpret $\widehat{\la}$ as the lump's position, and 
$\lambda$ as parametrizing its sharpness. The symmetric hedgehog has
$\lambda=0$, and large $\lambda$ lumps have taller, narrower energy density 
peaks than small $\lambda$ lumps.

One way of attacking the geodesic problem is to reduce its dimension by
identifying totally geodesic submanifolds, that is, choosing initial value
problems whose solution is simplified by some symmetry. The easiest method for
identifying such submanifolds is to find fixed point sets of discrete groups of
isometries. Any isometry maps geodesics to geodesics, so if there were a geodesic
starting off in the fixed point set of the isometry and subsequently deviating 
from it, this would be mapped under the isometry to another geodesic, identical
to the first throughout its length in the fixed point set, but deviating from
the set in a different direction. This violates the uniqueness of solutions of
ordinary differential equations, so no such geodesic may exist. If the initial
data are a point on the fixed point set and a velocity tangential to it, then the
geodesic must remain on the fixed point set for all subsequent time.

Examining (\ref{eq:g}) we see that
$\la\rightarrow-\la$ is an isometry. Its fixed point set is $S^{3}/\mbox{\set Z}_{2}$,
the isorotation orbit of the symmetric hedgehog, on which
the metric is
\begin{equation}
g=\frac{\pi}{6}\s\cdot\s.
\end{equation}
The kinetic energy is the rotational energy of a totally symmetric rigid body,
moment of inertia $\pi/6$. The solutions are just isorotations of the symmetric
hedgehog at constant frequency about some fixed axis. In this case isorotation 
is equivalent to spatial rotation because $\la=0\Rightarrow H=\mbox{\set I}$.

A less trivial geodesic submanifold is the fixed point set of the parity 
transformation described above, $M\mapsto\bar{M}$. This is a 3-dimensional
manifold, the product of the plane $\lambda_{2}=0$ in $\mbox{\set R}^{3}$ with
the circle $\{\alpha\in[0,\pi], \beta=\gamma=0\}\cup\{\alpha\in[0,\pi], \beta=\gamma=\pi\}$
in $S^{3}/\mbox{\set Z}_{2}$. The circle is more conveniently parametrized if we temporarily allow
$\alpha$ the domain $[0,2\pi]$, for it is then $\{\alpha\in[0,2\pi], 
\beta=\gamma=0\}$. This space contains lumps of arbitrary sharpness located on a 
great circle through the poles of \Ssp, each lump having an internal
phase, so certain of its geodesics may be candidates for ``travelling lumps.''
Introducing spherical polar coordinates for \la,
\begin{equation}
\la=\lambda(\sin\theta\cos\phi,\sin\theta\sin\phi,\cos\theta),
\end{equation}
the plane $\lambda_{2}=0$ is parametrized by $(\lambda,\theta)$ where $\theta
\in[0,2\pi]$, again gluing two semicircles together and extending the domain of
$\theta$ to cover the whole circle in one go. The metric on this geodesic 
submanifold is
\begin{equation}
g=(A+\lambda^{2}B)d\lambda^{2}+\lambda^{2}A\, d\theta^{2}+C\, d\alpha^{2}
-\lambda^{2}A\, d\theta\, d\alpha.
\end{equation}
So the kinetic energy is
\begin{equation}
T=\frac{1}{2}\left[(A+\lambda^{2}B)\dot{\lambda}^{2}+\frac{p_{\theta}^{2}C}{\lambda^{2}A(C-\lambda^{2}A/4)}+\frac{p_{\alpha}(p_{\alpha}+p_{\theta})}{C-\lambda^{2}A/4}\right]
\end{equation}
where we have used the cyclicity of $\theta$ and $\alpha$ to eliminate 
$\dot{\theta}$ and $\dot{\alpha}$ in favour of their constant, canonically
conjugate momenta,
\begin{eqnarray}
\label{eq:pdefs}
p_{\theta} &=& \lambda^{2}A\left(\dot{\theta}-\frac{1}{2}\dot{\alpha}\right) \\ \nonumber
p_{\alpha} &=& C\dot{\alpha}-\frac{1}{2}\lambda^{2}A\dot{\theta}.
\end{eqnarray}
Note that constant $p_{\theta}$ ($p_{\alpha}$) does {\em not} imply constant
$\dot{\theta}$ ($\dot{\alpha}$), nor does $p_{\theta}=0$ ($p_{\alpha}=0$) imply
$\dot{\theta}=0$ ($\dot{\alpha}=0$).

This system can be visualized as a point particle of position dependent mass
$(A+\lambda^{2}B)$ moving in a potential. It is the form of the potential  which
determines the broad qualitative features of its behaviour:
\begin{equation}
{\cal V}(\lambda)=p_{\theta}^{2}{\cal V}_{\theta}(\lambda)+
p_{\alpha}(p_{\alpha}+p_{\theta}){\cal V}_{\alpha}(\lambda).
\end{equation}
As can be seen from figure 2 while ${\cal V}_{\theta}(\lambda)$ is monotonically
decreasing, ${\cal V}_{\alpha}(\lambda)$ is monotonically increasing. This
allows the possibility of potential minima where the forces $-p_{\theta}^{2}
{\cal V}_{\theta}'(\lambda)$ outwards (in the sense of increasing $\lambda$) and
$p_{\alpha}(p_{\alpha}+p_{\theta}){\cal V}_{\alpha}'(\lambda)$ inwards are in
stable equilibrium. It certainly is {\em not} possible if 
$p_{\alpha}(p_{\alpha}+p_{\theta})\leq 0$, for then ${\cal V}(\lambda)$ as a 
whole is monotonically decreasing. This region in the $(p_{\alpha},p_{\theta})$
plane  is shown shaded in figure 3. Whatever the initial conditions on $\lambda$,
the lump always moves towards infinite $\lambda$ without passing through
$\lambda=0$ (which would correspond to the lump swapping hemispheres), reaching
the singularity $\lambda=\infty$, an infinitely tall, sharp spike, in finite
time. Thus $(\M_{1},g)$ is geodesically incomplete. This result follows from the 
rapid vanishing of the inertia to sharpening, $A+\lambda^{2}B$, in the large 
$\lambda$ limit (see figure 4). For example, consider the simple case 
$p_{\alpha}=p_{\theta}=0$ and let $\lambda(0)$ and $\dot{\lambda}(0)$ be 
strictly positive. It is easily seen that $t_{\infty}$, the time taken to reach 
the singular spike is proportional to the following integral:
\begin{equation}
t_{\infty}\propto \int_{\lambda(0)}^{\infty}\, d\lambda\, 
\sqrt{A(\lambda)+\lambda^{2}B(\lambda)}.
\end{equation}
The integrand is finite over the integration range (even if $\lambda(0)=0$), so
if $t_{\infty}$ diverges it can only be due to the large $\lambda$ behaviour.
But the integrand vanishes like $(\log\lambda)/\lambda^{2}$ at large $\lambda$,
fast enough to ensure convergence. The inclusion of repulsive potentials can only
make matters worse, so this singular behaviour extends to the rest of the shaded
area.

\vbox{
\centerline{\epsfysize=3truein
\epsfbox[70 190 550 590]{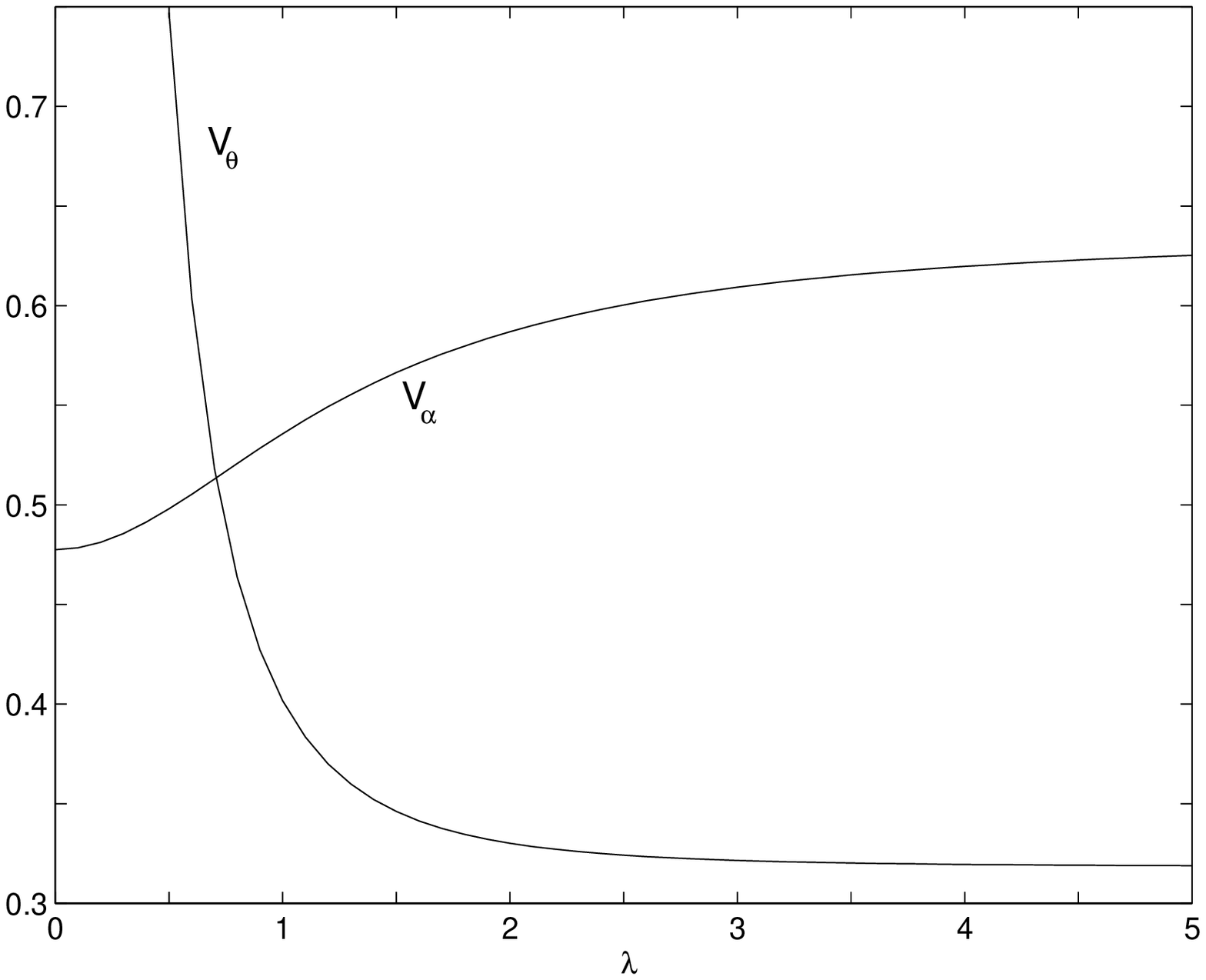}}
\centerline{\it Figure 2: The potential functions ${\cal V}_{\theta}(\lambda)$
and ${\cal V}_{\alpha}(\lambda)$.}
}
\newpage

\vbox{
\centerline{\epsfysize=2truein
\epsfbox[100 -400 490 -30]{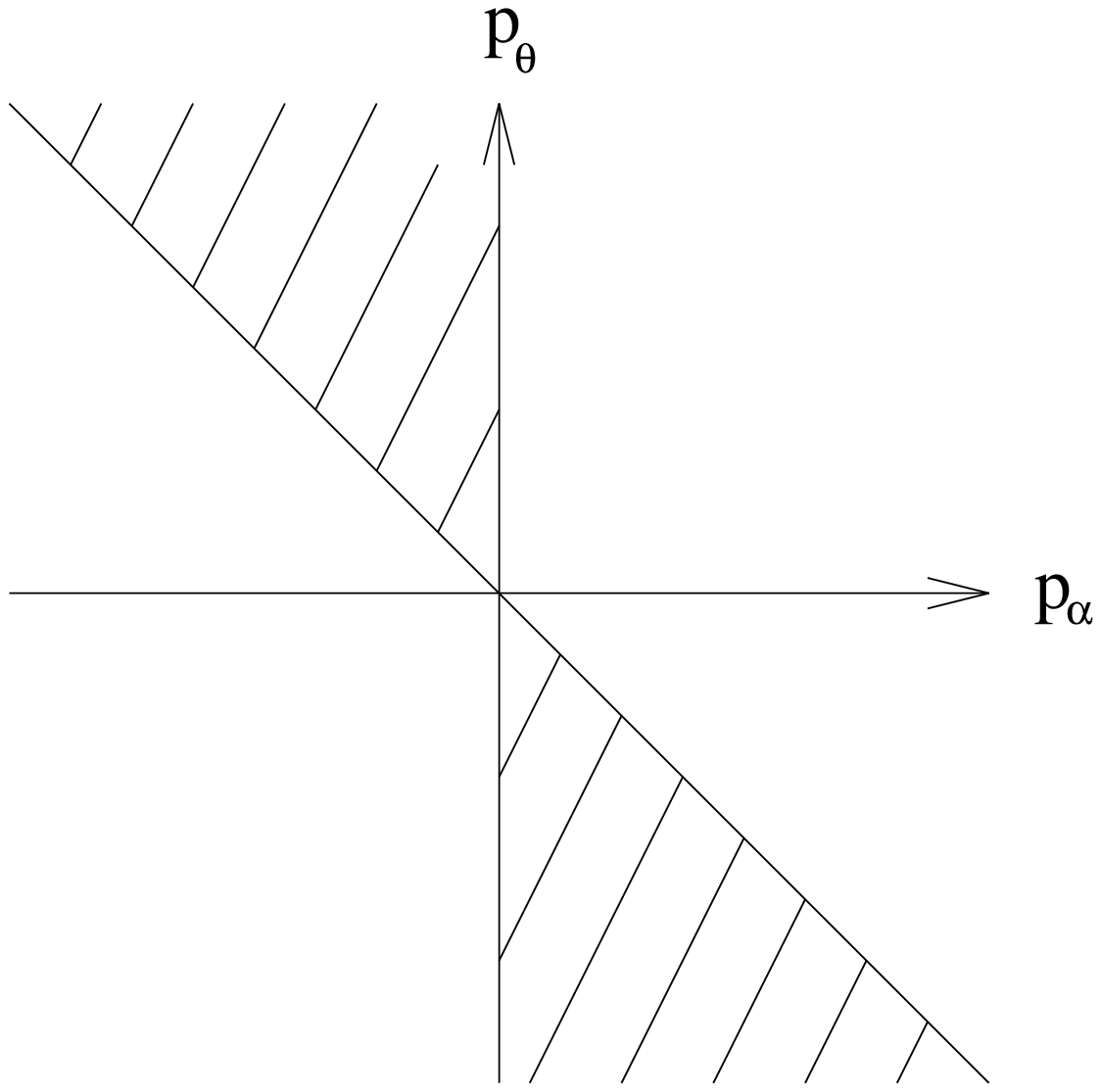}}
\centerline{\it Figure 3: The $(p_{\alpha},p_{\theta})$ plane. The instability
region is shaded and includes boundaries.}
}
\vspace{0.5cm}

\vbox{
\centerline{\epsfysize=3truein
\epsfbox[70 190 550 590]{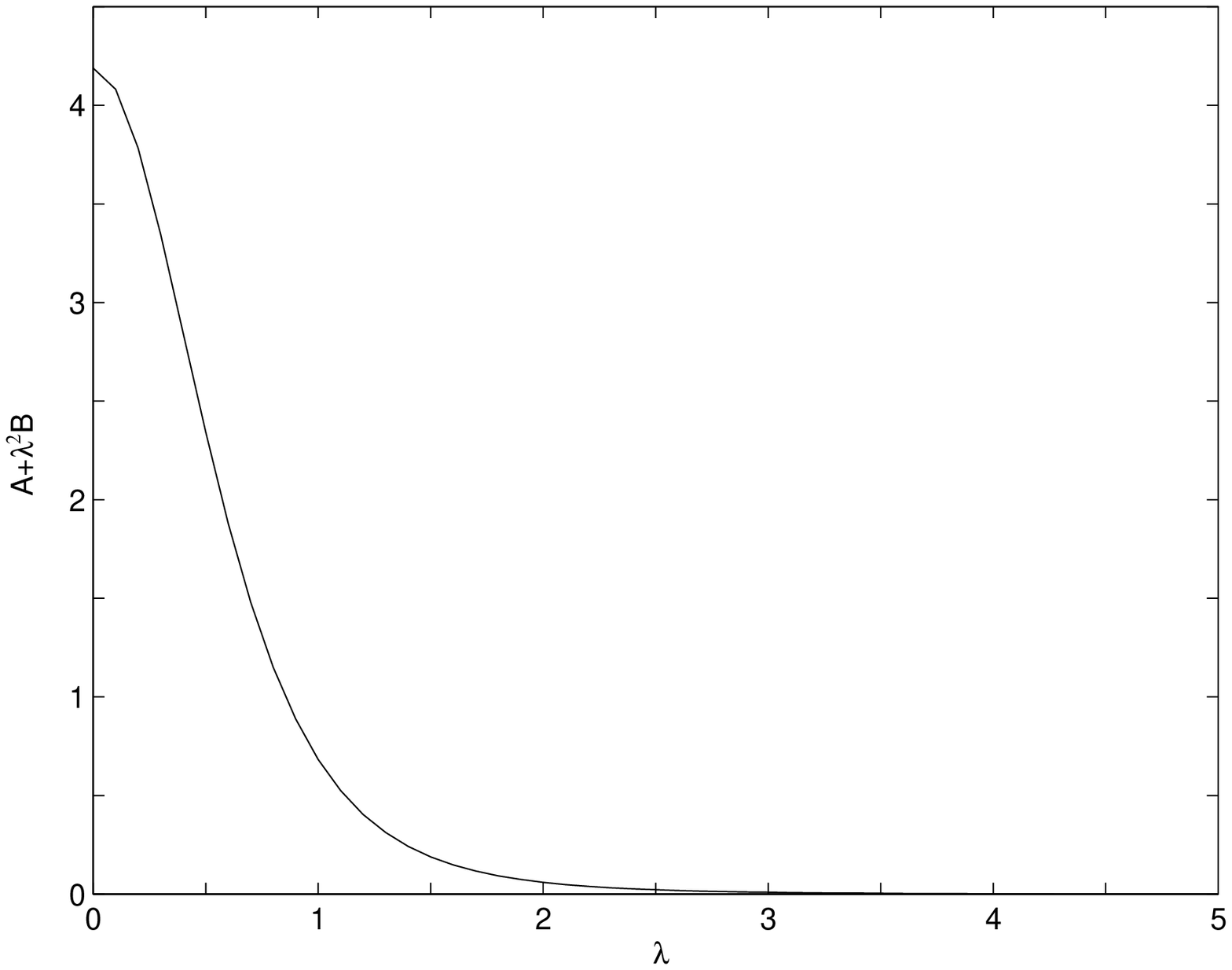}}
\centerline{\it Figure 4: The inertia to sharpening, $A+\lambda^{2}B$.}
}
\vspace{0.5cm}

In the unshaded region, one can define the positive constant $\kappa=
p_{\theta}^{2}/(p_{\alpha}^{2}+p_{\alpha}p_{\theta})$ such that
\begin{equation}
{\cal V}(\lambda)=p_{\alpha}(p_{\alpha}+p_{\theta})\left(
\kappa{\cal V}_{\theta}(\lambda)+{\cal V}_{\alpha}(\lambda)\right).
\end{equation}
Then the forms of the functions $-{\cal V}_{\theta}'$ and ${\cal V}_{\alpha}'$
(see figure 5) suggest that for each $\lambda$, there is one (and only one) value
of $\kappa$ (call it $\widetilde{\kappa}$) for which ${\cal V}$ has a minimum at
$\lambda$. The equilibrium condition is ${\cal V}'(\lambda)=0$, so
\begin{equation}
\widetilde{\kappa}(\lambda)=-\frac{{\cal V}'_{\alpha}(\lambda)}{{\cal V}'_{\theta}(\lambda)}.
\end{equation}
Inverting the definition of $\kappa$ we find that there are two distinct values 
of $p_{\alpha}/p_{\theta}$ for each $\widetilde{\kappa}$. If
$p_{\alpha}/p_{\theta}$ takes one of these and $\dot{\lambda}(0)=0$ then $\lambda$
will not subsequently change and hence $\dot{\alpha}$ and $\dot{\theta}$ will
also remain constant, allowing the lump to travel around a great circle on \Ssp\
with constant speed and shape while undergoing constant frequency isorotation.
The two values are
\begin{equation}
\nu_{\pm}(\lambda)
\equiv -\frac{1}{2}\pm\sqrt{\frac{1}{4}+\frac{1}{\widetilde{\kappa}(\lambda)}}.
\end{equation}
Substituting (\ref{eq:pdefs}) we can find the corresponding pair of stable
ratios $\dot{\alpha}/\dot{\theta}$ as functions of $\lambda$,
\begin{equation}
\omega_{\pm}\equiv
\frac{\lambda^{2}A(\lambda)\left(2\nu_{\pm}(\lambda)+1\right)}{
      2C(\lambda)+\lambda^{2}A(\lambda)\nu_{\pm}(\lambda)},
\end{equation}
(see figure 6). Thus, for any lump sharpness $\lambda$ and travel speed 
$\dot{\theta}$ there are two possible isorotation frequencies $\dot{\alpha}$
which allow stable, uniform travel and these two stability ``branches'' never
coincide.
It is interesting to note that $\lim_{\lambda\rightarrow\infty}\omega_{+}(\lambda)=1$
meaning that very tall, sharp lumps can travel uniformly with 
$\dot{\alpha}\approx\dot{\theta}$. Motion with constant $\lambda$ and 
$\dot{\alpha}=\dot{\theta}$ is simply constant speed spatial rotation carrying 
the lump around a great circle. So when the extent of the lump's structure is
negligible relative to the radius of curvature of \Ssp, it can travel in
analogous fashion to a flat-space \CP\ lump \cite{Ward}.

\vbox{
\centerline{\epsfysize=3truein
\epsfbox[70 190 550 590]{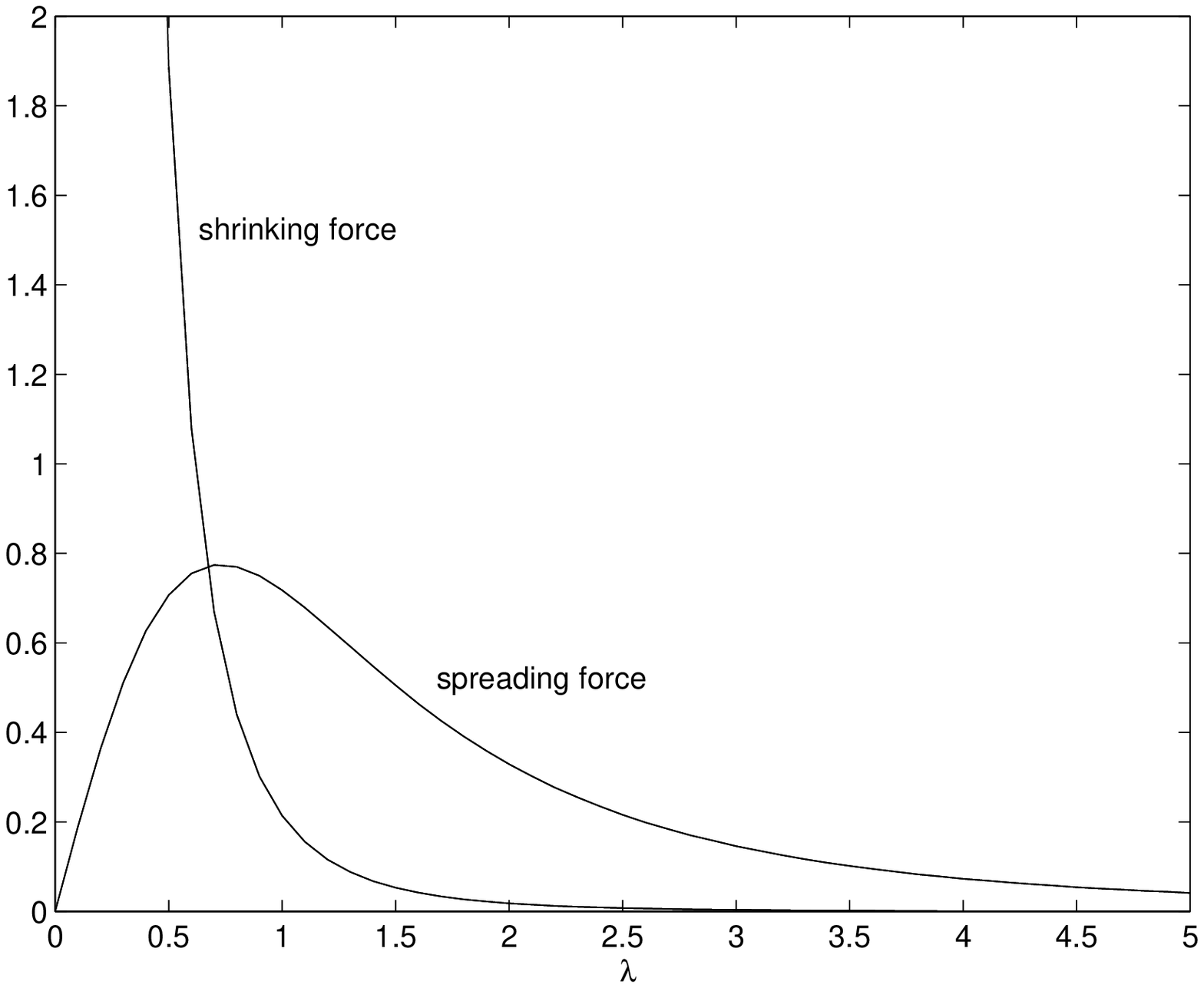}}
\noindent{\it Figure 5: The outward (shrinking) force, $-{\cal V}'_{\theta}(\lambda)$ and the inward (spreading) force, ${\cal V}'_{\alpha}(\lambda)$. The vertical scales of the two curves are different.}
}
\vspace{0.5cm}

\vbox{
\centerline{\epsfysize=3truein
\epsfbox[40 200 580 640]{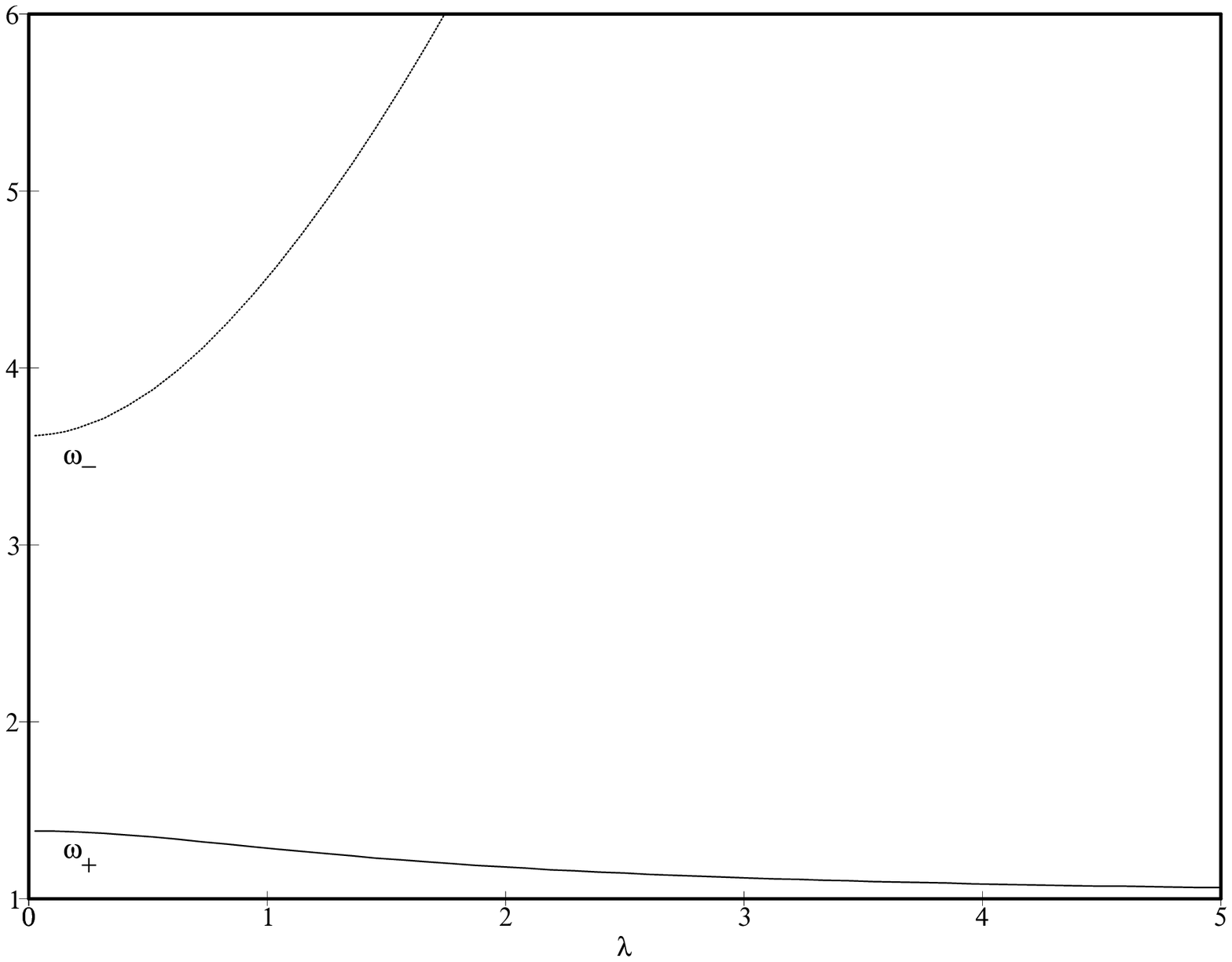}}
\centerline{\it Figure 6: The stable frequency ratios, $\dot{\alpha}/\dot{\theta}=\omega_{\pm}(\lambda)$. The upper curve $\omega_{-}(\lambda)$ tends to infinity at large $\lambda$.}
}
\vspace{0.5cm}

Since $\widetilde{\kappa}(\lambda)$ takes all positive values, whatever value
$\kappa$ takes there is an equilibrium $\lambda$. If $\lambda(0)$ is near this 
value, then (assuming $|\dot{\lambda}(0)|$ is not too large) the 
shape of the lump will oscillate periodically about the preferred sharpness, and
its speed of travel round the sphere will vary with the same period. If 
$|\dot{\lambda}(0)|$ {\em is} too large, or the lump is initially much too 
spread out for its $\kappa$, then it will escape to the singular spike in finite
time.

Let us examine the concrete example $p_{\theta}=p_{\alpha}$. Figure 7 shows 
the potential ${\cal V}$ with its minimum and the lump travel speed $\dot{\theta}$
as functions of $\lambda$. We imagine a particle of position dependent mass
moving in this potential and for simplicity take $\dot{\lambda}(0)=0$. Clearly,
if  we release the particle with $\lambda(0)<\lambda_{A}\approx 0.626$, it will
move off to infinity, the last case mentioned above. But if $\lambda(0)>
\lambda_{A}$, oscillatory motion ensues. Even here there are two qualitatively
different cases, because $\dot{\theta}(\lambda)$ has an absolute minimum at
$\lambda_{C}\approx 2.096$, a turning point which is only reached if 
$\lambda(0)<\lambda_{B}\approx 0.789$ (or $\lambda(0)>\lambda_{C}$). If
$\lambda_{B}<\lambda(0)<\lambda_{C}$ then the speed of travel oscillates in 
simple phase with the lump sharpness, going from fast, spread-out lump to slow,
sharp lump and back again. But if $\lambda_{A}<\lambda(0)<\lambda_{B}$ or
$\lambda(0)>\lambda_{C}$ the speed undergoes an extra wobble during the middle of the
sharpness cycle, speeding up then slowing down again as it passes through its
maximum sharpness. This case corresponds to lumps whose shape oscillates more
acutely.

\vbox{
\centerline{\epsfysize=3truein
\epsfbox[70 190 550 590]{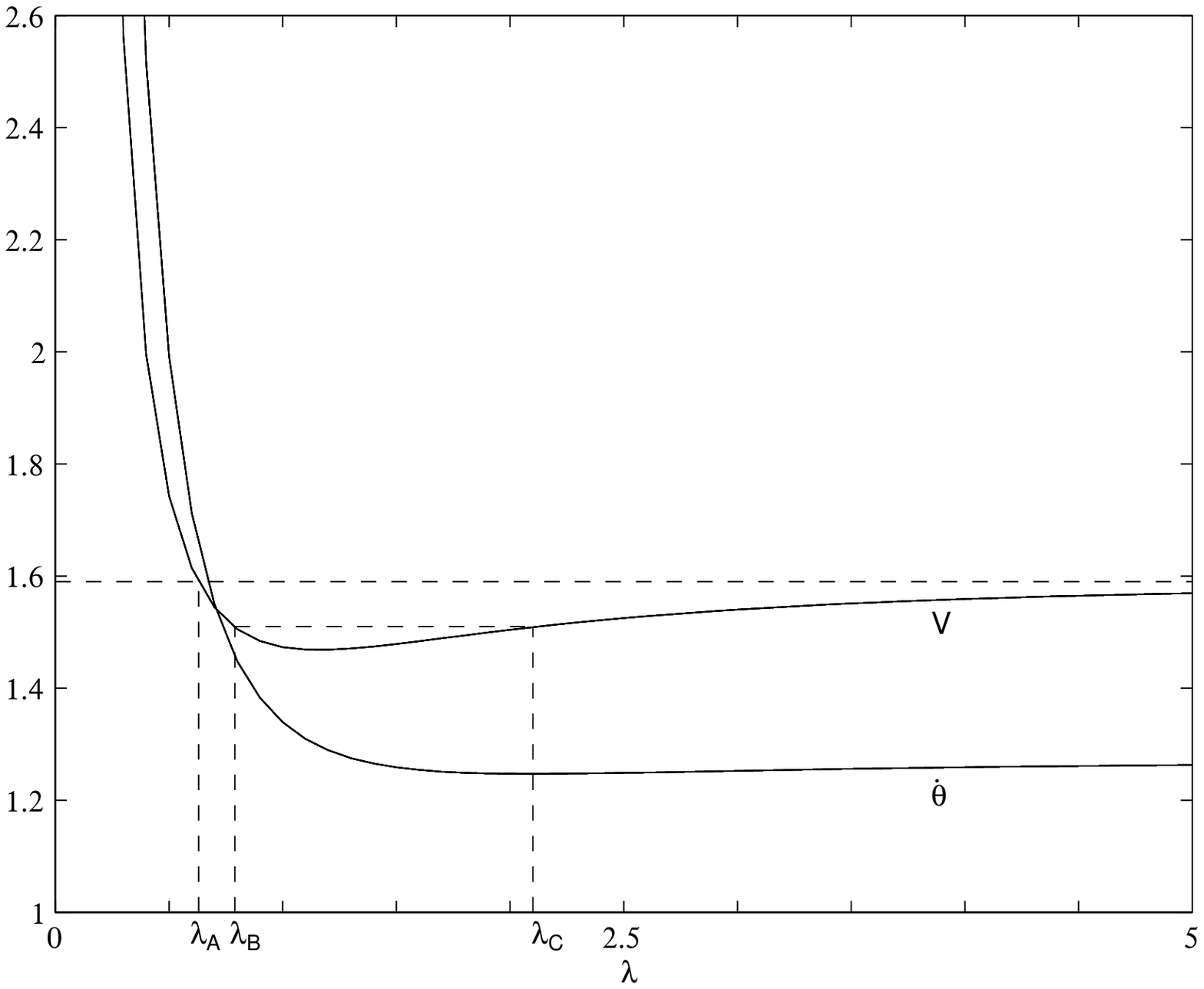}}
\centerline{\it Figure 1: The $p_{\theta}=p_{\alpha}$ case: potential ${\cal V}$ and speed $\dot{\theta}$.}
}
\vspace{0.5cm}

Other interesting geodesic submanifolds are generated by computing the fixed
point sets $\Sigma_{ab}$ of the isometries 
$M\mapsto(i\tau_{a})^{\dagger}M(i\tau_{b})$, simultaneous isorotation and spatial
rotation by $\pi$ about the $a$ and $b$ axes respectively:
\begin{eqnarray}
U &\mapsto& (i\tau_{a})^{\dagger}U(i\tau_{b}), \nonumber \\
H &\mapsto& (i\tau_{b})^{\dagger}H(i\tau_{b}) \nonumber \\
\Rightarrow \lambda_{c} &\mapsto&
\left\{\begin{array}{rl} \lambda_{c} & c=b \\
                        -\lambda_{c} & c\neq b.
                        \end{array} \right.
\end{eqnarray}
Thus if $q\in \Sigma_{ab}$ \la\ must point along the $b$-axis. On $\Sigma_{bb}$,
$U=\exp(i\psi\tau_{b}/2)$ where $\psi\in[0,2\pi]$, whereas if $a=b\pm 1 \bmod 3$
then $U=\exp(\pm i\pi\tau_{c}/4)\exp(i\psi\tau_{b}/2)$ where $c=b\mp 1 \bmod 3$.
It follows that \s\ also points along the $b$-axis, independent of $a$. The
$a\neq b$ submanifolds are the images of $\Sigma_{bb}$ under $\pm\pi/2$
isorotations about the three axes, so it suffices to solve the geodesic problem
on $\Sigma_{bb}$ -- geodesics on $\Sigma_{ab}, a\neq b$, are then obtained by
acting with the appropriate isometry. The choice of $b$ doesn't matter,
and we choose to study the cylinder $\Sigma_{33}\cong S^{1}\times\mbox{\set R}$
consisting of lumps of every sharpness located at the North (South) pole if
$\lambda_{3}>0$ ($\lambda_{3}<0$), arbitrarily rotated about the North-South
axis. Note that $[U,H]=0$ on $\Sigma_{33}$ so ``isorotated'' and ``spatially
rotated'' mean the same thing in this case.

The kinetic energy on $\Sigma_{33}$ is
\begin{equation}
T=\frac{1}{2}\left[(A+\lambda_{3}^{2}B)\dot{\lambda}_{3}^{2}+
\frac{p_{\psi}^{2}}{C+\lambda_{3}^{2}D}\right]
\end{equation}
where once again $p_{\psi}$ is the momentum conjugate to $\psi$,
\begin{equation}
p_{\psi}=(C+\lambda_{3}^{2}D)\dot{\psi},
\end{equation}
and is constant by virtue of the cyclicity of $\psi$. This looks like a
particle in one dimension moving in a potential
\begin{equation}
p_{\psi}^{2}{\cal V}_{\psi}=\frac{p_{\psi}^{2}}{2(C+\lambda_{3}^{2}D)}
=\frac{1}{2}p_{\psi}\dot{\psi}
\end{equation}
with postion dependent mass. From the potential (figure 8) we see that all
motion is oscillatory and that $\lambda_{3}$ periodically changes sign. Thus a
lump set spinning about its own axis will spread out, its rotation slowing,
until it is uniformly spread over the sphere, whereupon it will shrink to its
mirror image in the opposite hemisphere, regaining its original spin speed as it
does so. The process then reverses and the lump ``bounces'' between antipodal 
points indefinitely.

\vbox{
\centerline{\epsfysize=3truein
\epsfbox[70 190 550 590]{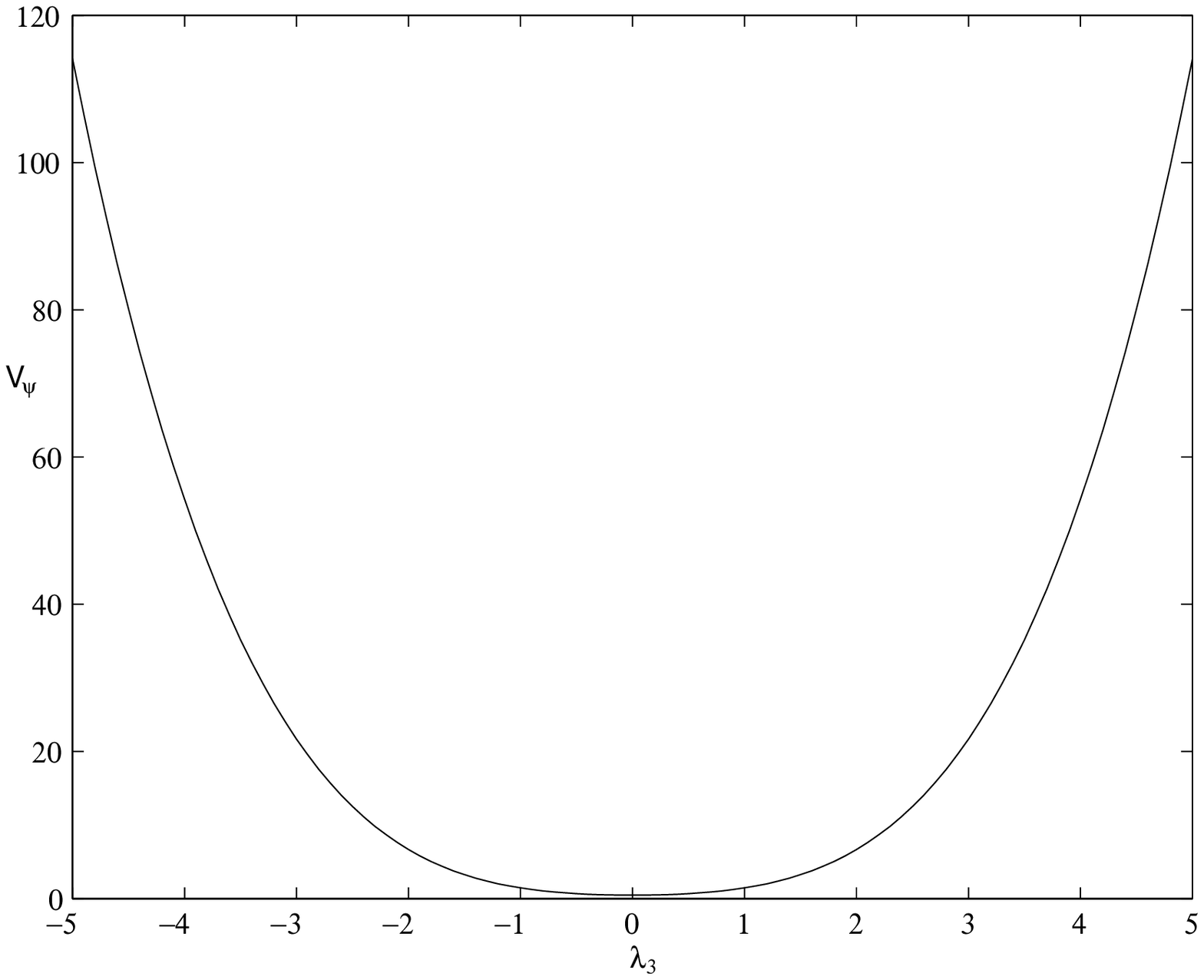}}
\centerline{\it Figure 8: Potential, ${\cal V}_{\psi}(\lambda_{3})$.}
}
\vspace{0.5cm}

Defining the new coordinate
\begin{equation}
s(\lambda_{3})=\int_{0}^{\lambda_{3}}\, d\mu\, \sqrt{A(\mu)+\mu^{2}B(\mu)},
\end{equation}
which takes values in a finite open interval $(-\rho,\rho)$ symmetric about $s=0$,
the metric on $\Sigma_{33}$ becomes
\begin{equation}
g=ds^{2}+H(s)d\psi^{2}
\end{equation}
where $H\left(s(\lambda_{3})\right)=C(\lambda_{3})+\lambda_{3}^{2}D(\lambda_{3})$.
Since $|dH/ds|<1\; \forall s$, the manifold may be embedded as a surface of 
revolution in $\mbox{\set R}^{3}$ and geodesics on it can be visualized 
directly. Figure 9 is a sketch of the embedded surface, which is of finite length
and sausage-shaped with its ends pinched to infinitely sharp spikes, the tips
of which are the points $\lambda_{3}=\pm\infty$ and so are missing. The 
coordinates $(s,\psi)$ are geodesic orthogonal coordinates: a curve of constant
$\psi$ is a geodesic along the length of the cylinder, lying in a plane
containing the cylinder's axis, parametrized by arc length $s$, while a curve of
constant $s$ is a circle of radius $H(s)$, lying in a plane orthogonal to the
cylinder's axis. Two such curves always intersect at right angles.

The spinning geodesics described above wind around the cylinder, never reaching
the ends (this would violate conservation of ``angular momentum'') but winding
back and forth between two circles $s=\pm\widetilde{\rho}$, 
$\widetilde{\rho}<\rho$ which they touch tangentially. The angle $\varphi$ at 
which the geodesic intersects the circle $s=0$ determines 
$\widetilde{\rho}$. When $\varphi=0$ the 
geodesic stays on the circle, $\widetilde{\rho}=0$ (a spinning symmetric
hedgehog) and $\widetilde{\rho}(\varphi)$ monotonically increases, tending to the
supremum $\rho$ as $\varphi$ tends to $\pi/2$ ($\varphi=\pi/2$ is an irrotational 
geodesic between antipodal singular spikes). Note that the geodesic 
incompleteness already mentioned appears again, this time characterized by the 
finite length of the cylinder and the missing points $s=\pm\rho$ 
($\lambda_{3}=\pm\infty$).

\newpage

\vbox{
\centerline{\epsfysize=4truein
\epsfbox[250 220 400 620]{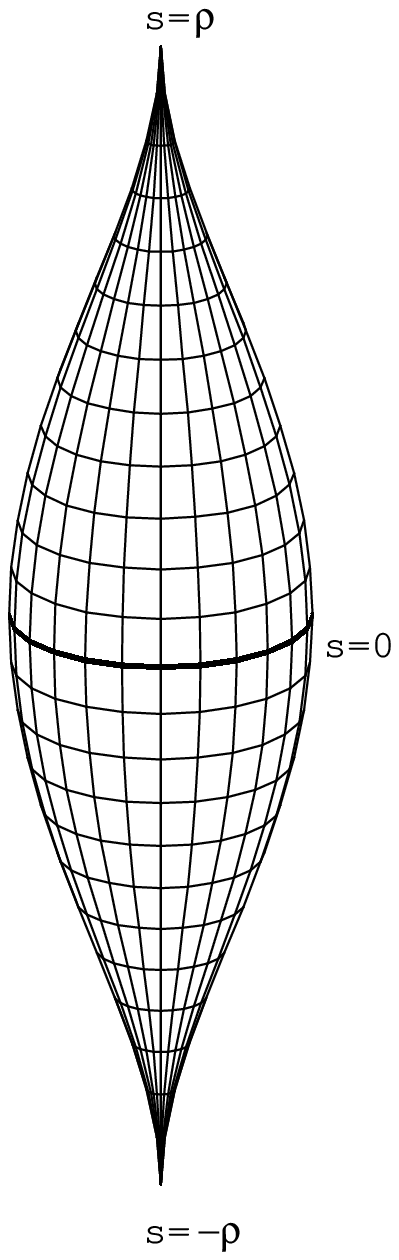}}
\centerline{\it Figure 9: A sketch of the geodesic submanifold $\Sigma_{33}$ embedded as a surface of revolution in $\mbox{\set R}^{3}$.}
}
\vspace{0.5cm}

\section{Concluding remarks}

The behaviour of isolated topological solitons in flat space is generally rather 
trivial, whereas, as we have seen, despite the homogeneity of $S^{2}$, the motion of
a single lump on the sphere is surprisingly complicated. It does travel on great
circles, but while doing so its shape may oscillate in phase with its speed,
whose periodic variation is of one of two types depending on the violence of the
shape oscillations, or it may collapse to an infinitely tall, thin spike in
finite time. A lump sent spinning about its own axis spreads out then re-forms in
the opposite hemisphere, endlessly commuting between antipodal points.

The infinities in the unit-charge metric in flat space can be attributed to the
lumps' polynomial tail-off: the kinetic energy needed to rigidly spin or 
scale-deform a lump diverges because such motions involve changing the field at
spatial infinity. The \CP model on any compact space should be free of this 
problem because the kinetic energy, being an integral over a space of finite
volume, must be finite provided the kinetic energy density is non-singular. 
Conversely, one would expect the singularity to persist in the model defined on
hyperbolic space.

The flat-space \CP model can be made more ``physical'' by adding a $(2+1)$-dimensional
version of the Skyrme term to stabilize against lump collapse, and a potential to
stabilize against spread. The Bogomol'nyi bound remains valid but unsaturable.
The potential is somewhat arbitrary, but one interesting possibilty
\cite{Piette} gives a mass to small amplitude travelling waves
of the \p\ field, termed pions in analogy with the Skyrme model, and gives the
lump an exponential rather than polynomial tail. This allows the lump to
rotate, a problem if one attempts a collective coordinate approximation to 
low-energy dynamics along the lines recently proposed in \cite{Shah,Stuart}. The
idea is to restrict the field to the ``Bogomo'nyi regime'' moduli space (in this
the space of static \CP solutions), introducing a potential and a perturbed
metric (in \cite{Shah} but not \cite{Stuart}) to account for the new interactions, which are 
assumed to be weak. There seems little hope of perturbing the singular flat-space
metric such that rotations become possible, but the problem does not arise on
the sphere. 

The geodesic approximation could be used to investigate the interaction of two
lumps moving on $S^{2}$.
Right angle scattering in head on collisions emerges naturally from the
geodesic approximation of many flat-space models as a consequence of the
classical indistinguishability of topological solitons. It would be interesting 
to see if there is some analogous behaviour on the sphere. However, evaluating
the two-lump metric could be difficult since the action of the isometry group 
on the charge-2 moduli space is far less accessible than in the present case.
Even in flat space \cite{Leese}, the scattering problem is sufficiently 
complicated to
require considerable numerical effort.

\vspace{0.5cm}
\noindent
{\bf Acknowledgments:} I would like to thank Richard Ward, who suggested this
work, and Bernd Schroers for many useful discussions. I also acknowledge the
financial support of the UK Particle Physics and Astronomy Research Council.


\begin{thebibliography}{xx}

\bibitem{Din} A M Din and W J Zakrzewski,
``General classical solutions in the $\mbox{\set C}P^{N-1}$ model''
{\sl Nucl. Phys.} {\bf B174} (1980) 397

\bibitem{Ward} R S Ward,
``Slowly moving lumps in the \CP model in $(2+1)$ dimensions''
{\sl Phys. Lett.} {\bf 158B} (1985) 424

\bibitem{Leese} R A Leese,
``Low-energy scattering of solitons in the \CP model''
{\sl Nucl. Phys.} {\bf B344} (1990) 33

\bibitem{Zak} R A Leese, M Peyrard and W J Zakrzewski,
``Soliton stability in the $O(3)$ $\sigma$-model in $(2+1)$ dimensions''
{\sl Nonlinearity} {\bf 3} (1990) 387

\bibitem{Woo} G Woo,
``Pseudoparticle configurations in two-dimensional ferromagnets''
{\sl J. Math. Phys.} {\bf 18} (1977) 1264

\bibitem{Belavin} A A Belavin and A M Polyakov,
``Metastable states of two-dimensional isotropic ferromagnets''
{\sl JETP Lett.} {\bf 22} (1975) 245

\bibitem{Nakahara} M Nakahara,
{\sl Geometry, Topology and Physics}
(Adam Hilger, Bristol, England, 1990), chapter 4

\bibitem{Rajaraman} R Rajaraman,
{\sl Solitons and Instantons}
(North Holland, Amsterdam, The Netherlands, 1989), pp 48--58

\bibitem{Schwer} H Schwerdtfeger
{\sl Geometry of Complex Numbers}
(Dover, New York, USA, 1979), p 41

\bibitem{Penrose} R Penrose and W Rindler
{\sl Spinors and space-time} Volume 1
(Cambridge University Press, Cambridge, England, 1984), pp 14--21

\bibitem{Manton} N S Manton,
``A remark on the scattering of BPS monopoles''
{\sl Phys. Lett.} {\bf 110B} (1982) 54

\bibitem{Ryan} M P Ryan and L C Shepley,
{\sl Homogeneous Relativistic Cosmologies}
(Princeton University Press, Princeton, USA, 1975), chapter 6

\bibitem{Choquet} Y Choquet-Bruhat, C DeWitt-Morette and M Dillard-Bleick,
{\sl Analysis, Manifolds and Physics} Part I
(North-Holland, Amsterdam, The Netherlands, 1982), p 184

\bibitem{Piette} B M A G Piette, B J Schroers and W J Zakrzewski,
``Multi-solitons in a two-dimensional Skyrme model''
Durham University preprint DTP 94-23, to appear in {\sl Zeitschrift f\"{u}r
Physik C}

\bibitem{Shah} P A Shah,
``Vortex scattering at near-critical coupling''
Cambridge University preprint DAMTP 94-8, to appear in {\sl Nucl. Phys. B}

\bibitem{Stuart} D Stuart,
``Dynamics of abelian Higgs vortices in the near Bogomolny regime''
{\sl Commun. Math. Phys.} {\bf 159} (1994) 51

\end{thebibliography}
\end{document}